\documentclass[twoside]{article}
\usepackage{fleqn,espcrc2}
\usepackage{epsfig}

\newcommand{\AmS}{{\protect\the\textfont2
  A\kern-.1667em\lower.5ex\hbox{M}\kern-.125emS}}

\title{Highlights of CP 2000}

\author{John Ellis\address{Theoretical Physics Division, CERN \\ 
        CH 1211 Geneva 23}}%
        
\def\beq{\begin{equation}}
\def\eeq{\end{equation}}
\def\bea{\begin{eqnarray}}
\def\eea{\end{eqnarray}}
\def\bq{\begin{quote}}
\def\eq{\end{quote}}

\def\gappeq{\mathrel{\rlap {\raise.5ex\hbox{$>$}}
{\lower.5ex\hbox{$\sim$}}}}

\def\lappeq{\mathrel{\rlap{\raise.5ex\hbox{$<$}}
{\lower.5ex\hbox{$\sim$}}}}

\def\Toprel#1\over#2{\mathrel{\mathop{#2}\limits^{#1}}}

\begin{document}
\begin {abstract}
Various developing topics in CP violation are reviewed. There are many
theoretical reasons to hope that the CKM paradigm may be incomplete.  It
is surely too soon to be claiming new physics in
$\epsilon^\prime/\epsilon$ or in $D^0-\bar D^0$ mixing, but rare $K$
decays offer interesting places to search for new physics. It is probably
also premature to see a clash between global CKM fits and current
estimates of $\sin \beta$ and $\gamma$, where much more precise data will
soon be available. There are interesting possibilities to look for CP
violation in neutrino oscillations and in Higgs physics. Rapid progress
can be expected now that CP violation is moving to the top of the particle
physics agenda. 
\\
%{~~~~~~~~~}\\
\begin{center}
{CERN-TH/2000-297~~~~~~~~~~~~~~hep-ph/0011396}
\end{center}
\vspace{1pc}
\end{abstract}

\maketitle

\section{Preface}

My personal interest in CP violation is driven by the search for physics
beyond the Standard Model, and this bias guides the following remarks. You
all know that the Standard Model is sadly deficient. It is theoretically
very unsatisfactory, despite agreeing with all confirmed accelerator data.
The Standard Model does not explain the particle quantum numbers (Q, I, Y,
Colour) and contains 19 arbitrary parameters. Three of these are its
independent gauge couplings, and there is a {\it CP-violating strong
interaction vacuum angle} $\theta_{QCD}$, to which we return later. The
arbitrary Yukawa couplings are linked to the 6 quark masses, 3
charged-lepton masses, 3 weak mixing angles and the {\it CP-violating
Kobayashi-Maskawa phase}. Finally, there are the $W^\pm$ and Higgs boson
masses:  is the latter about 115 GeV~\cite{heroes}?

The first clear evidence for physics beyond the Standard Model has come
from neutrino oscillations~\cite{nosc}, which are most plausibly
interpreted in terms
of 3 non-zero neutrino masses and 3 neutrino mixing angles. In addition,
there are expected to be 3 {\it CP-violating neutrino phases}, one of
which is observable in scattering experiments, with the other two playing
a potential r\^ole in neutrinoless double-$\beta$ decay. There may well be
additional parameters associated with neutrino mass generation, e.g., for
more Higgs fields and/or heavy singlet neutrinos.

Additional parameters beyond the Standard Model appear when one considers
gravity and cosmology, such as Newton's constant, the cosmological
`constant' and at least one parameter to characterize inflation. Moreover,
at least {\it one additional CP-violating parameter} is needed to explain
the baryon asymmetry of the Universe, which cannot be explained by any of
the parameters mentioned above~\cite{bbb}. 

As already mentioned, we know that some physics beyond the Standard Model
is needed to explain neutrino oscillations and, we hope, link them to
other physics. The Higgs boson might be one doorway to new physics: if it
weighs 115 GeV, there must be some new physics at an energy $\lappeq 10^6$
GeV~\cite{EGNO}, to prevent the Higgs potential from becoming unstable at
this scale~\cite{AI}.
To my mind, the default option for this new physics is supersymmetry, as
discussed shortly. 

What are the prospects for observable physics beyond the Standard Model in
CP-violating quantities, such as
$\epsilon^\prime/\epsilon$~\cite{epsprime}~\footnote{For the 
first non-zero estimate of $\epsilon^\prime/\epsilon$ in the 
Standard Model, see~\cite{EGN}. Paradoxically, the confirmation that
$\epsilon^\prime/\epsilon\not= 0$ has killed one candidate for new
physics, namely the superweak model of CP violation in the $K^0$
system~\cite{Sweak}.},
$D^0-\bar D^0$ mixing~\cite{DDbar} or $\sin
2\beta$~\cite{CDF,Babar,Belle}? As discussed later, it is really
too soon to interpret any of these as evidence for new physics. On the
other hand, there are some intriguing prospects for finding physics beyond
the standard Model in rare $K$ and/or $B$ decays.

As emphasized here by Masiero~\cite{Masiero}, supersymmetry is a framework
for physics beyond the Standard Model that is well motivated by the
naturalness (or hierarchy)  problem of electroweak symmetry breaking. This
suggests that sparticles weigh $\lappeq$ 1 TeV, making supersymmetry a
testable hypothesis that is already tightly constrained by experiments, in
particular on CP violation~\cite{Kane}. To simplify, let us assume $R$
conservation.  Then, if the soft supersymmetry-breaking scalar masses
$m_0$ and gaugino masses $m_{1/2}$ are assumed to be universal, and also
the trilinear $A$ terms, there are 5 real parameters and 2 new
CP-violating phases. These may be taken as the phases of $A$ and
$m_{1/2}$, in this `minimal' supersymmetric model (often called the CMSSM
or mSUGRA). In this case, the CP-violating asymmetry in $b\rightarrow
s\gamma$ decay may be a promising signature. If one relaxes the
universality assumption, there are 60 real parameters and 43 CP-violating
phases. If one makes the `reasonable' assumption that all the
universality-breaking parameters are of the same order as the
corresponding quark and lepton masses: $\Delta \tilde m = {\cal O}(m_q)$,
then there may also be supersymmetric signatures in $\epsilon_K,
\epsilon^\prime/\epsilon$ and $K^0_L\rightarrow\pi^0e^+e^-$.  One may even
ask the question whether all the observed CP violation, i.e., both
$\epsilon_K$ and $\epsilon^\prime/\epsilon$, could originate from
supersymmetry~\cite{all}. Finally, one might be `hopeful' and assume no
close connection between flavour breaking for conventional and
supersymmetric particles. In this case, there could be many experimental
signatures, including most notably $d_n, \mu\rightarrow e\gamma$ and
$B\rightarrow K_S(\phi$ or $\pi$). 

\section{Light Quarks: $u, d, s$}

As we heard here from Hinds~\cite{Hinds}, the experimental upper limit on
the neutron electric dipole moment is $d_n < 6.3 \times 10^{-26}$ e.cm.
This is far above the Standard Model prediction $d_n\sim 10^{-31}$ e.cm,
providing an extensive opportunity to search for new physics. In
particular, supersymmetric models often predict large values of $d_n$,
which has often been quoted as constraining new supersymmetric phases
$\phi$ to the percent level. However, there are many possibilities for
cancellations among different diagrams, e.g., charginos vs
gluinos~\cite{Kane2} and/or diagrams involving $d$ and $s$
quarks~\cite{EF}. The latter are present in the neutron wave function at
the 10 \% level, and often have supersymmetric amplitudes that are larger
by a factor $m_s/m_d\sim 20$. Thus one should be cautious before inferring
very strong constraints on supersymmetric phases. It should also not be
forgotten that $d_n$ is uniquely sensitive to $\theta_{QCD}$, constraining
it to be $\lappeq 10^{-9}$: a subject to which we return later.

The electric dipole moments of other systems are also interesting. For
example, that of $^{199}$Hg is sensitive to CP-violating nucleon-nucleon
interactions, though these are difficult to calculate reliably. The dipole
moment of the electron is also interesting, though there may still be
cancellations between symmetric diagrams. The electric dipole moment of
the muon is expected in many models to be larger than that of the electron
by a factor $m_\mu/m_e \sim 200$, and might provide an interesting
opportunity for the BNL ($g-2)_\mu$ ring, or for future muon storage
rings. 

The pioneering CP-violating observable $\epsilon_K$ is difficult to
calculate reliably in the Standard Model. Certainly, its measured value is
qualitatively consistent with the CKM prediction, but it is
subject to uncertainties in non-perturbative
hadronic matrix elements. Therefore, to my mind it is better
not to use
$\epsilon_K$ in global fits, and I would be difficult to convince myself
of any signature for new physics in $\epsilon_K$. 

As already mentioned, it is ironic that the confirmation of a non-zero
value for $\epsilon^\prime/\epsilon$ killed one possible example of
physics beyond the Standard Model, namely the superweak
theory~\cite{Sweak}. However,
any conclusions from the value of $\epsilon^\prime/\epsilon$ may be
premature before its experimental value has settled down further. Once
again, it is predicted qualitatively correctly by the Standard Model.
However, $\epsilon^\prime/\epsilon$ is particularly difficult to calculate
exactly, because of large cancellations between contributions associated
with different non-perturbative matrix elements. For
this reason, it will
also be quite hard to convince oneself of any need for new physics in
$\epsilon^\prime/\epsilon$, until one has a better handle on these
hadronic matrix elements~\footnote{For a recent discussion of the
importance of final-state $\pi \pi$ interactions, see~\cite{PP}.}. They
have been tackled
using a variety of phenomenological approaches, including the chiral quark
model~\cite{epsprimeth1} and the
$1/N_C$ expansion~\cite{epsprimeth2}. The best answers may eventually be
supplied by lattice
calculations~\cite{epsprimelatt}, but these are not yet very precise.

Rare $K$ decays~\cite{rareK} provide the best light-quark opportunity to
look for new
physics.  The Standard Model calculations are relatively reliable, and far
below the present experimental upper limits~\footnote{Except for
$K^+\rightarrow\pi^+\bar\nu\nu$~\cite{Kpinunu}.}.  New physics might
appear
in a
flavour-changing $Z$ vertex, $Z_{sd}$, in an anomalous dipole moment for
$\gamma$ or gluon emission, ${1\over M} \bar q \sigma^{\mu\nu} f_{\mu\nu}
q$, or in some four-fermion operator, ${1\over M^2} (\bar ff)(\bar
ff)$~\cite{Masiero,Silvestrini}.
The best prospects may be offered by $Z_{sd}$, which is unsuppressed by a
potentially large mass $M$. In contrast to 
$K^+\rightarrow\pi^+ \bar\nu\nu$, the Standard Model predictions
for the rare decay
$K^0_L\rightarrow\pi^0\bar\nu\nu$ and the direct contribution to the
$K^0_L\rightarrow\pi^0e^+e^-$ amplitude are well below the experimental
measurement limits. Somewhat larger values are possible in
supersymmetric models with minimal flavour violation, i.e., the same
pattern as in
the Standard Model, and considerably larger values in a more general class
of supersymmetric models with
arbitrary flavour violation, or with a general parametrization of the
$Z_{sd}$
vertex, constrained by $\epsilon_K, \epsilon^\prime/\epsilon$ and
$K^0_L\rightarrow\mu^+\mu^-$~\cite{Masiero}. 

There is an active outgoing programme on $K^+\rightarrow\pi^+\bar\nu\nu$
decay~\cite{futurekpinunu}, led currently by the BNL E787$\rightarrow$E949
experiment, aiming
at a sensitivity of 5 to 10 events within the Standard Model.  There is
also a proposal at FNAL to gather $\sim 100~
K^0\rightarrow\pi^0\bar\nu\nu$ decays, and proposals at both BNL and FNAL
to gather up to $\sim 100~K^0_L\rightarrow\pi^0\bar\nu\nu$ decays. In view
of the interest of these decays for flavour physics in general, and CP
violation in particular, I hope the resources will be made available to
conduct such next-generation rare $K$ decay experiments, particularly if
one recalls the amount of resources directed towards $B$ physics. 

\section{Medium-Heavy Quark: $c$}

One expects $D^0-\bar D^0$ mixing to be very small in the Standard Model:
$x \equiv \Delta m / r \lappeq$ 0.001 and $y \equiv \Delta\Gamma/\Gamma$
smaller still, providing another opportunity for new physics to strut its
stuff.  There is currently some excitement~\cite{BGLNP} generated by the
FOCUS report
of a $2-\sigma$ signal for $y \sim 0.03$~\cite{DDbar}.  This excitement is
probably
premature, since the effect is not confirmed. It is important to clarify
what is the maximum possible value in the Standard Model, which may be
rather larger than some estimates, through presumably not as large as the
central value of FOCUS. There is also the issue of compatibility between
the FOCUS result and that of CLEO~\cite{CLEO}, which would be easier if
there is a
large strong-interaction phase.  Large CP violation in the $D^0-\bar D^0$
system would be a particularly strong indication for physics beyond the
Standard Model. 

\section{Global CKM fits}

\begin{figure}
\epsfig{figure=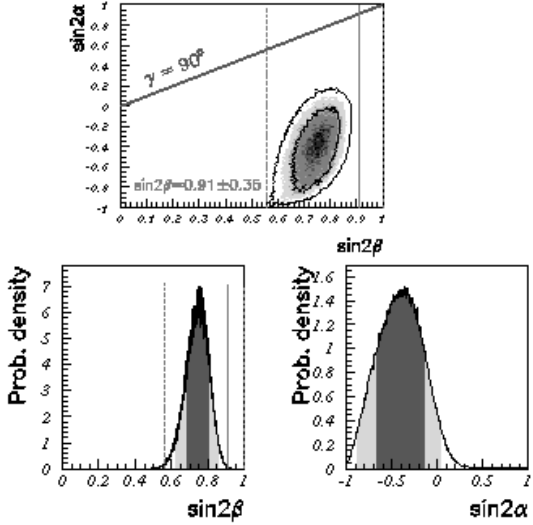,width=7.5cm}
\caption{\it Results of a global CKM fit, displaying
preferred ranges of $\sin 2 \beta$ and $\alpha$, and
indicating that $\gamma < 90^0$~\cite{CKMfit}.} 
\label{fig:Parodi}
\end{figure}

Before discussing the search for CP violation in $B$ meson decays, let us
review the Standard Model predictions provided by global fits of the CKM
parameters~\cite{CKMfit}.  Several convergent measurements indicate that
\bea
\bar\rho &=& 0.206 \pm 0.043~, \nonumber \\
\bar\eta &=& 0.339 \pm 0.044
\label{one}
\eea
including the $B^0-\bar B^0$ mass difference $\Delta m_d$, the lower limit
on the $B^0_s-\bar B^0_s$ mass difference $\Delta m_s$~\cite{HFWG}, and
the ratio
$\vert V_{ub}\vert/\vert V_{cb}\vert$. The ranges (\ref{one}) are
supported by the experimental value of $\epsilon_K$, though the
theoretical uncertainties in its interpretation are rather greater. As
seen in Fig.~\ref{fig:Parodi}, the
corresponding prediction for $\beta(=\phi_1)$ is: 
\beq
\sin 2 \beta = 0.723 \pm 0.069
\label{two}
\eeq
and it would be quite surprising if experiments find $\sin 2 \beta < 0.5$, as
discussed later. 
On the other hand, the value of $\alpha$ is rather more
uncertain:
\beq
\sin 2 \alpha = -0.28 \pm 0.27
\label{three}
\eeq
The global Standard Model fit predicts quite confidently that $\gamma$ is in the
first quadrant:
\beq
\gamma = 58.5 \pm 6.9^0 ~, \quad
< \, 90^0 \, @ \, 99.8 \% ~{\rm c.l.}
\label{four}
\eeq
This result is mainly enforced by the $\Delta m_s$ constraint, and it is worth
remembering that a joint analysis of the existing experiments yields a 2.5 -
$\sigma$ effect corresponding to $\Delta m_s \sim 18/ps$~\cite{HFWG}, as
expected in the
Standard Model on the basis of the other CKM measurements.

\section{Heavy Quark: $b$}

With the advent of the B factories, a new age is dawning for CP violation
studies. However, there is still some early-morning fog!

\noindent
\underline{$\sin 2 \beta$}\\
The present experimental measurements:
\bea
\sin 2 \beta &=& 0.79 \pm 0.39 \pm 0.16 ~~~\cite{CDF} \nonumber \\
&& 0.12 \pm 0.37 \pm 0.09 ~~~ \cite{Babar} \nonumber \\
&& 0.45~^{+ 0.43 + 0.07}_{-0.44 - 0.09} ~~~~~~~~~ \cite{Belle}
\label{five}
\eea
are mutually compatible within their large errors, and may be combined to yield
\beq
\sin 2 \beta = 0.42 \pm 0.24
\label{six}
\eeq
This value is also compatible at the $1-\sigma$ level with the global CKM
fit prediction (\ref{two}).  However, this has not prevented theorists
from speculating on the possible consequences if it eventually transpires
that $\sin 2 \beta < 0.5$~\cite{mad}. 

Could we have misunderstood the Standard Model so badly that $\sin 2 \beta
< 0.5$ is possible within the conventional CKM framework? This would
require reducing the estimated value of $\vert V_{ub}/V_{cb}\vert$ by
about 30\%, or increasing the estimate of $\xi \equiv \sqrt{{B_{B_s}\over
B_{B_d}}}~~{f_{B_s}\over f_{B_d}}$ to $\gappeq 1.4$, or increasing $\hat
B_K$ to $\gappeq 1.3$, or some combination of these effects. Among these,
a different value of $\vert V_{ub}/V_{cb}\vert$ may be the least
implausible~\cite{Nir}. 

Alternatively, there might be new physics in $B^0 - \bar B^0$ mixing,
which could yield $\sin2 \beta < 0.5$ if the new physics has a non-trivial
phase: $2\beta\rightarrow 2(\beta + \theta_d)$, or if it changes the
magnitudes of $\Delta m_{d,s}$ by factors $\Gamma_{d,s} : 0.5 \lappeq
\Gamma_d \lappeq 1$ and $\Gamma_s/\Gamma_d \gappeq 1.1$. Another
possibility is new physics in $K^0-\bar K^0$ mixing, which would alter the
interpretation of $\epsilon^\prime$~\cite{Nir}.

However, I emphasize again that there is no need for new physics to explain the
value of $\sin 2 \beta$ (yet). We await with impatience the next round of
analyses!

\noindent
\underline{$\sin 2 \alpha$}\\
The prospects for measuring this quantity in $B_0 \rightarrow \pi^+\pi^-$ decay
may now be less problematic than in 1999, for two reasons. One is that the two
new measurements of the branching ratio:
\bea
B (B_0&\rightarrow& \pi^+\pi^-) = \nonumber \\
&&\left\{
\matrix{(9.3^{+2.6+1.2)}_{(-2.3 -1.4}) \times 10^{-6}~\cite{Babarpipi}
 \cr \cr
(6.3 \pm 4.0) \times 10^{-6}~\cite{Bellepipi}\hfill}\right.
\label{seven}
\eea
raise the possibility that this decay mode may be less suppressed relative
to $K^+\pi^-$ than was suggested by the previous CLEO
data~\cite{CLEO}. This would
have the twin advantages of making it easier to extract from the
background and of reducing the prospective penguin pollution. The second
ray of hope for $\sin 2 \alpha$ is that the systematic perturbative QCD
factorization approach~\cite{factn}, based on transition form factors with
calculable
corrections, should permit better calculations of penguin pollution from
first principles~\cite{VLB}. 

\begin{figure}
\epsfig{figure=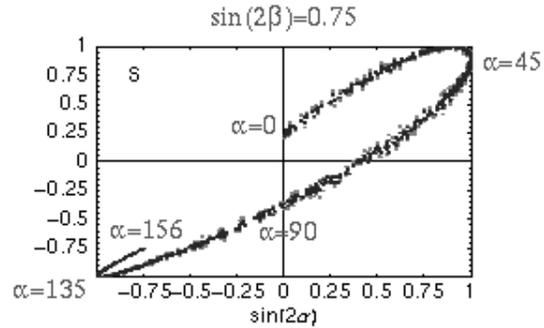,width=7.5cm}
\caption{\it The correlation between $S$ and $\alpha$
expected in the QCD factorization approach to penguins~\cite{VLB}.}
\label{fig:Neubert2}
\end{figure}

Recall that the CP-violating asymmetry in $\stackrel{(-)}{B}^0 \rightarrow
\pi^+\pi^-$ decay can be written as
\beq
A_{CP}(t) = -S\cdot \sin (\Delta m_dt) + C\cdot \cos (\Delta m_dt)
\label{eight}
\eeq
where $S = \sin 2\alpha$ and $C = 0$ in the absence of penguins. In the
presence of calculable penguins, a tight correlation between the value of
$\alpha$ and $S$ can be predicted, and the prediction for $S$ checked by
verifying the corresponding prediction for the direct CP-violating
quantity $C$~\cite{VLB}, as seen in Fig.~\ref{fig:Neubert2}. 

\noindent
\underline{$\gamma$}\\
There has recently been intense theoretical exploration of the bounds on
$\gamma$ that can be set indirectly by measurements of ratios of branching
ratios for $B \rightarrow K\pi$ and $\pi\pi$ decays~\cite{gammabounds}. In
particular, there
has been some excited comment that perhaps $\gamma > 90^0$, based on the
early (and imprecise)  measurements of these branching ratios. If true,
this might be an interesting hint of physics beyond the Standard Model,
which strongly prefers $\gamma < 90^0$ as discussed earlier. 
\begin{figure}
\epsfig{figure=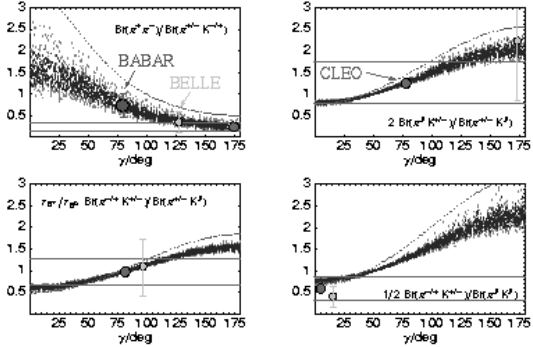,width=7.5cm}
\caption{\it Measurements of exclusive $B$ decays by CLEO, BaBar
and Belle do not yet permit any conclusion whether $\gamma <$ 
or $> 90^0$~\cite{VLB}.} \label{fig:Neubert}
\end{figure}
However, any
talk of a $\gamma$ `crisis' is premature, in view of the various sources
of theoretical uncertainty (renormalization scale, quark masses, wave
functions, $\ldots$) and the large experimental errors (so far). In
particular, what will happen to the ratio of $\pi^+\pi^-/K^+\pi^-$~? Let
us wait for an experimental consensus to emerge. 

\section{Very Heavy Quark: $t$}

The production and decays of $\bar tt$ pairs will be an interesting place
to look for CP violation in the future~\cite{ttbar}, but was not discussed
at this
meeting, so I do not discuss it further here. 

\section{Neutrino Masses and Oscillations}

There is no good reason why $m_\nu= 0$, since vanishing masses are
generally associated with exact gauge symmetries (e.g., the photon and the
$U(1)$ gauge invariance of QED), and there is no massless gauge boson
coupling to lepton number. A neutrino mass term $m_\nu~\nu \cdot \nu$ is
possible if there is a $\Delta L = 2$ interaction, and is a generic
feature of GUTs. However, it is even possible in the Standard Model, $
without$ new particles or gauge interactions, if one simply introduces a
non-renormalizable interaction~\cite{BEG}: 
\beq
{1\over M}~ (\nu H)\cdot (\nu H) \Rightarrow m_\nu = 
{<0\vert H \vert 0>^2\over M}
\label{nine}
\eeq
scaled by some new heavy mass scale $M \gg m_W$ so that $m_\nu \ll
m_{q,\ell}$.  In
GUTs, one normally encounters `right-handed' singlet neutrinos $\nu_R$,
and a seesaw mass matrix of the generic form
\beq
(\nu_L, \nu_R)~~\left(\matrix{0 & m\cr  m^T & M}\right) ~~\left( \matrix{\nu_L
\cr \nu_R}\right)
\label{ten}
\eeq
where the off-diagonal entry $m = {\cal O} (m_{q,\ell})$, yielding
\beq
m_\nu = m~{1\over M}~ m^T
\label{eleven}
\eeq
after diagonalization.

The quantities $m, M$ in (\ref{ten}), (\ref{eleven})
should be understood as 3$\times$3 matrices, that cannot in general be expected
to be diagonalizable in the same basis as the charged-lepton masses
$m_\ell$,
leading to a neutrino mixing matrix \`a la CKM~\cite{MNS}:
\begin{equation}
V_{MNS} \; = \; V_{L} V_\nu^\dagger
\label{MNS}
\end{equation}
with an observable CKM-like phase $\delta$. However, in addition, there
are also relative phases in the diagonalized light Majorana neutrino mass
matrix $diag(e^{i\alpha}, e^{i\beta},1)$. These are unobservable at
energies $E\ll m_{\nu_i}$, but enter into the interpretation of
double-$\beta$ decay experiments. 

The mixing angles $\theta_{ij}$ and the CP-violating phase $\delta$
appearing in (\ref{MNS}) are observable via neutrino oscillations, and
there are indications of non-zero values~\cite{nosc}.  In view of the
different origin (\ref{ten},\ref{eleven}) for neutrino masses, it should
perhaps not be surprising if neutrino mixing angles are very different
from the CKM angles. Indeed, atmospheric neutrino experiments favour large
mixing between $\nu_\mu$ and $\nu_\tau$~\cite{atmosc}, and solar neutrino
experiments favour (less strongly) large mixing between $\nu_e$ and some
combination of $\nu_\mu$ and $\nu_\tau$~\cite{solarosc}.  On the other
hand, atmospheric and reactor neutrino experiments have so far only
established upper limits on $\nu_\mu\rightarrow\nu_e$ mixing.
 
The next stage of neutrino mixing studies will be led by long-baseline
accelerator beam experiments~\cite{LBnu}, which should measure $\nu_\mu$
mixing more precisely, set better limits on (or measure)
$\nu_\mu\rightarrow \nu_e$ mixing, and observe $\tau$ production by
$\nu_\mu\rightarrow\nu_\tau$ oscillations.

The following stage may be led by neutrino factories -- muon storage rings
that produce intense neutrino beams with known fluxes, spectra and
charge/flavour separation~\cite{Gomez}. Such neutrino factories consist of
an intense
proton source, typically providing several MW of protons at some energy
between 2 and 20 GeV, a high-power target producing pions, a system to
capture, cool and accelerate the muons produced by pion decays, and then a
`ring' to store the muons while they decay. The ring need not be circular,
indeed it should have long straight sections (that need not be parallel)
producing $\nu$ beams directed towards detectors at different distances. 

Such a neutrino factory would further refine measurements of neutrino
mixing angles, but the prime interest would be to look for CP
violation~\cite{CPnu} via the asymmetry
\beq
A_{CP} \equiv {P(\nu_\mu\rightarrow\nu_e) -
P(\bar\nu_\mu\rightarrow\bar\nu_e)\over
P(\nu_\mu\rightarrow\nu_e) + P(\bar\nu_\mu \rightarrow \bar\nu_e)}
\label{thirteen}
\eeq
which has the following approximate theoretical expression:
\beq
A_{CP} \simeq 2 c_{13} s_{13} s_{23} \sin \delta {\Delta m_{12}^2 L \over
2 E_\nu} \sin^2 {\Delta m_{23}^2 L \over 4 E_\nu}
\label{forteen}
\eeq
where $\Delta m^2_{12}$ is the smaller neutrino mass-squared difference
measured in solar neutrino experiments. The latest Super-Kamiokande
results favour relatively large $\Delta m^2_{12} \gappeq 2\times 10^{-5}$
eV$^2$ and a large mixing angle, favouring the measurement of $A_{CP}$. 

\begin{figure}
\epsfig{figure=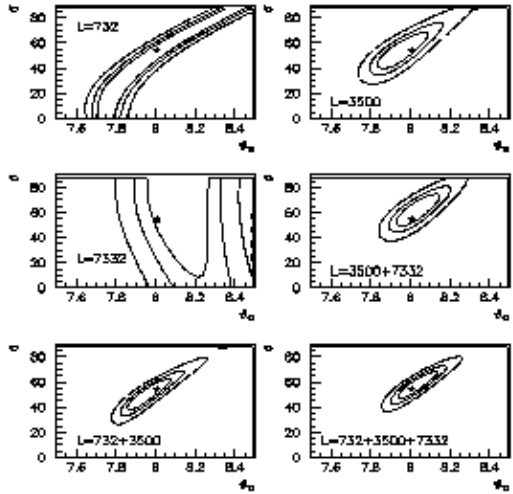,width=7.5cm}
\caption{\it Illustration how neutrino oscillation data
over different baselines may be used to measure $\theta_{13}$ and the
CP-violating phase $\delta$, assuming $10^21$ $\mu$ decays
and a realistic detector~\cite{Gomez}.} \label{fig:Gomez}
\end{figure}

The measurement of this quantity must contend with the CP-asymmetric
matter environment of the Earth. The best significance for measuring
$A_{CP}$ is likely to be obtained with a baseline $L \sim$ 3000
km~\cite{Gomez,CPnu}, e.g., CERN to Northern Scandinavia or the Canary
Islands. As seen in Fig.~\ref{fig:Gomez}, there would be good sensitivity
to the
CP-violating phase $\delta$ with $\gappeq 10^{21} \mu$ decays. 
Alternatively, it has been suggested that one could measure $\delta$ with
a low-energy $\nu_\mu$ beam produced directly ty $\pi$
decay~\cite{lowECP}, but this seems very difficult with the beam
insensitivites currently discussed, and suffers from non-trivial
backgrounds~\cite{Gomez}. 

\section{Higgs Bosons}

CP violation in the Higgs sector has long been a favoured alternative (or
supplement) to the CKM model. The minimal Standard Model
cannot
accommodate CP violation in the Higgs sector. In the MSSM, the Higgs
sector does not violate CP at the tree level, but may acquire CP violation
via radiative corrections~\cite{MSSMHiggsCP}.  Assuming universal scalar
masses $m_0$,
gaugino masses $m_{1/2}$ and trilinear supersymmetry-breaking parameters
A, there are two potentially-important sources of CP violation in the
MSSM, namely Arg($A_{t,b})$ and Arg$(m_{\tilde g})$. These induce Higgs
scalar-pseudoscalar mixing at the one- and two-loop level,
respectively~\cite{CEPW}: 
\beq
\delta~ M^2_{SP} 
\sim {m^4_t\over v^2}~~{\mu~{\rm Im} A_t \over 32 \pi^2 m^2_{susy}} +
\ldots
\label{fifteen}
\eeq
In the presence of CP violation, it is best to parametrize the MSSM Higgs
sector in terms of $m_{H^+}$ and $\tan\beta$, since the pseudoscalar Higgs
mass $m_A$, the conventional mass parameter choice, is no longer an
eigenvalue of the neutral Higgs mass matrix~\cite{CEPW2}.

We have proposed a new MSSM Higgs benchmark scenario (CPX) with maximal CP
violation~\cite{CEPW2}:
\bea
m_{\tilde q} &=& m_{\tilde t} = m_{\tilde b} \equiv m_{susy}~,~~~ \mu = 4
m_{susy}~,
\nonumber \\
\vert A_t\vert& = &\vert A_b\vert = 2m_{susy}~,~~~\vert m_{\tilde g}\vert = 
1~{\rm TeV} \nonumber \\
{\rm Arg} (A_t)& =& {\rm Arg} (m_{\tilde g}) = 90^9
\label{sixteen}
\eea
This we contrast with a CP-conserving scenario (MAX) with maximal mixing,
in which
\bea
A_t = A_b &=& \sqrt{6} m_{susy}~,\nonumber \\
\mu = m_{\tilde B} &=& m_{\tilde W} =
200~{\rm GeV} \nonumber \\
{\rm Arg} (A_t) &=& {\rm Arg} (m_{\tilde g}) = 0
\label{seventeen}
\eea
\begin{figure}
\epsfig{figure=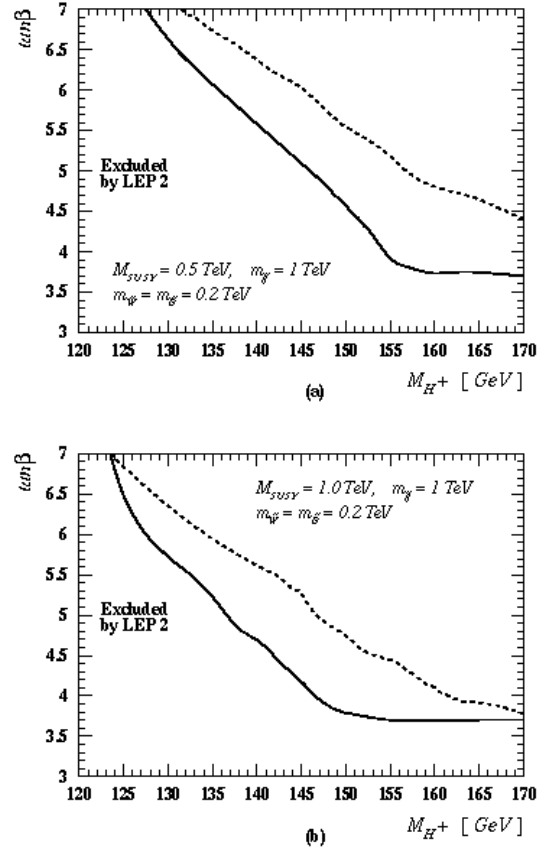,width=7.5cm}
\caption{\it Comparison between the regions of the
$(m_{H^+}, \tan \beta)$ plane allowed in the maximal
CP-violating scenario (\ref{sixteen}) and the CP-conserving 
scenario (\ref{seventeen}), for two different
supersymmetry-breaking scales~\cite{CEPW2}.} 
\label{fig:CPXMAX}
\end{figure}
The LEP physics reaches for the CPX and MAX scenarios are compared in
Fig.~\ref{fig:CPXMAX}~\cite{CEPW2}.

\begin{figure}
\epsfig{figure=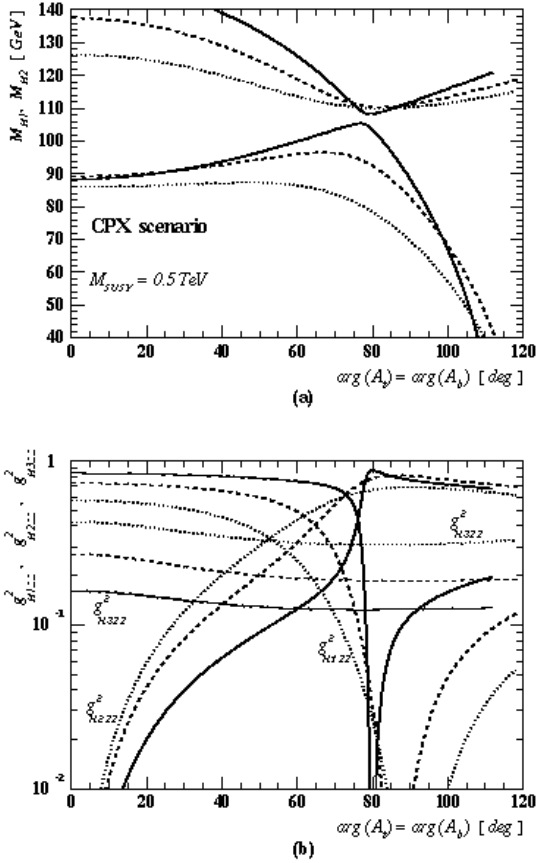,width=7.5cm}
\caption{\it Variation of the two lightest neutral Higgs masses
$M_{H1, H2}$ and the $Hi-Z-Z$ couplings, as functions of
the CP-violating angle Arg($A_{t,b})$. Note the level
crossing in the top panel and the vanishing $H1-Z-Z$
coupling in the lower panel~\cite{CEPW2}.} 
\label{fig:levelX}
\end{figure}

Among the intriguing possibilities offered by the CP-violating MSSM is level
crossing when the two lightest neutral Higgses have $m_{H_1}\simeq m_{H_2}$. As
seen in Fig.~\ref{fig:levelX}, this is accompanied by a
strongly-suppressed ZZ coupling:
\beq
g_{H_1ZZ}\vert_{CPX} \ll g_{H_1ZZ}\vert_{SM}
\label{eighteen}
\eeq
resulting in suppressed production of the lightest MSSM Higgs at LEP~2. On
the
other hand,
\beq
g_{H_1\bar bb}\vert_{CPX} \simeq g_{H_1\bar bb}\vert_{MAX}
\label{nineteen}
\eeq
and the dominant decay modes is expected to be $\bar bb$, as usual. 
In view of
(\ref{eighteen}), we find that an $H_1$
as light as 50 GeV might have escaped detection at LEP~\cite{CEPW2}.

An intriguing feature of the level-crossing scenario, also seen in
Fig.~\ref{fig:levelX},
is the fact that the second Higgs $H_2$ weighs $\sim$ 115 GeV. If the
current weak Higgs `signal' at LEP~\cite{heroes} turns out to be
confirmed, is it
conceivable that the experiments are seeing {\it the second MSSM Higgs
boson} ? It is also not excluded experimentally that there might exist
another Higgs
weighing $\sim$ 95 GeV with $g^2_{HZZ} \simeq
0.1~g^2_{HZZ}\vert_{SM}$,
which could be accommodated in the MSSM if $m_{H^+} \sim$ 160 GeV,
$\tan\beta\sim $, ${\rm Arg}(A_t) \sim 85^0$~\cite{CEPW2}. 

It would require an extensive programme of measurements at the LHC, an
$e^+e^-$ linear collider and a muon collider to unravel potential CP
violation in the MSSM Higgs sector. 

\section{Baryogenesis}

As already emphasized, the fact that we exist is evidence for physics
beyond the Standard Model. Generating a cosmological baryon asymmetry
requires a departure from thermal equilibrium as well as sufficient CP
violation~\cite{Sakharov,bbb}. In the Standard Model, the LEP lower limit
on
$m_H$ implies
that the electroweak phase transition is not first order~\cite{notfirst},
and, in
addition, the Standard Model has too little CP violation.
The following
are some of the favoured scenarios for baryogenesis. 

\noindent
\underline {GUT Boson Decay?}\\
This scenario might seem remote from experimental test, but many such scenarios
would induce a renormalization of the non-perturbative QCD vacuum parameter
$\theta_{QCD}$, and hence a neutron electric dipole moment that is close to the
present experimental upper limit~\cite{EGNR}:
\beq
d_n \gappeq 10^{-27} e.cm~,
\label{twenty}
\eeq
to be contrasted with the Standard Model value $d_n \sim 10^{-31}$ e.cm. Of
course, the interpretation of any observation of $d_n$ in the range
(\ref{twenty}) would be ambiguous, but at least it offers some prospect of
testing GUT baryogenesis scenarios.

\noindent
\underline{Heavy Neutrino Decay?}\\
A variation on the GUT boson decay scenario is the out-of-equilibrium
decay of a heavy `right-handed' neutrino seen in
(\ref{ten})~\cite{nuRdecay}. In this
case, the relevant CP violation would be associated with the Dirac
neutrino couplings appearing in the off-diagonal entries in (\ref{ten})
$\lambda = m/<0\vert H \vert 0>$. These generate a cosmological lepton
asymmetry, which is then partially converted by electroweak sphalerons
into a baryon asymmetry.  Could this scenario be tested by measuring CP
violation in $\nu$ oscillations? Unfortunately, they measure a different
form of CP violation, although there might well be some relation between
them.

\noindent
\underline{Supersymmetric Electroweak Transitions?}\\
In the MSSM, if there is a light stop $\tilde t$, as well as a relatively
light Higgs boson, it is possible for the electroweak phase transition to
be first order~\cite{notfirst,MSSMfirst}. Moreover, as we have seen, large
CP violation in the MSSM
cannot be excluded. Possible tests of this scenario include finding the
Higgs and/or a light $\tilde t$, as well as supersymmetric CP violation. 
This scenario is the only one to offer a prospect of relating the observed
value of $n_B/n_\gamma$ directly to quantities measurable in the
laboratory.

\section{(In-)Conclusions}

The experimental situation with regard to CP violation is evolving
rapidly: after $\epsilon_K$ and the confirmation that $\epsilon^\prime /
\epsilon \not= 0$, we can look forward soon to measurements of $\sin
2\beta, \sin 2 \alpha, \gamma$ and other CP-violating observables in B
decays. There are also prospects for observing CP violation elsewhere,
e.g., in electric dipole moments. In parallel, theoretical tools such as
the heavy-quark effective theory and factorization are also developing
rapidly. It is too soon to get excited about new physics in
$\epsilon^\prime/\epsilon$, $D^0-\bar D^0$ mixing, $\sin 2\beta$ or
$\gamma$, but we should be vigilant. CP violation is moving from the
qualitative era to a quantitative one, when deviations from the Standard
Model may well appear. In the longer run, new prospects for probing CP
violation are opening up, e.g., in the neutrino sector and perhaps even
the Higgs sector.

Generally, the flavour problem is now moving to the top of the particle
physics agenda, with the completion of precision electroweak measurements
at LEP and possibly the first element in the solution of the mass problem
with a hint of the Higgs boson~\cite{heroes}. Our excitement makes it
difficult to wait for the next conference in this series, but, while we
do, there are grounds for worry -- that the CKM model may be right -- and
also for hope -- that it may be wrong, or at least incomplete.

\end{document}